\documentclass[aps,prl,reprint]{revtex4-1}
\usepackage{graphicx}
\usepackage{color}
\usepackage{dcolumn}
\usepackage{bm}
\usepackage{ucs}
\usepackage{physics}
\usepackage{hyperref}
\usepackage{float}
\usepackage[modulo]{lineno}
\usepackage{tikz}

\begin{document}
\title{
Direct  Transport between 
Superconducting Subgap States in a Double Quantum Dot }
\author{G. O. Steffensen$^{1}$}
\author{J. C. Estrada Salda\~{n}a$^{1}$}
\author{A. Vekris$^{1,2}$}
\author{P. Krogstrup$^{1,3}$}
\author{K. Grove-Rasmussen$^{1}$}
\author{J. Nyg{\aa}rd$^{1}$}
\author{A. L. Yeyati$^{4}$}
\author{J. Paaske$^{1}$}

\affiliation{$^{1}$Center for Quantum Devices, Niels Bohr Institute, University of Copenhagen, 2100 Copenhagen, Denmark}
 \affiliation{$^{2}$Sino-Danish College (SDC), University of Chinese Academy of Sciences}
\affiliation{$^{3}$Microsoft Quantum Materials Lab Copenhagen, Niels Bohr Institute, University of Copenhagen, 2100 Copenhagen, Denmark}
\affiliation{$^{4}$Departamento de F\'isica Te\'orica de la Materia Condensada, Condensed Matter Physics Center (IFIMAC), and Instituto Nicol\'as Cabrera, Universidad Aut\'onoma de Madrid, 28049 Madrid, Spain}

\begin{abstract}
We demonstrate direct transport between two opposing sets of Yu-Shiba-Rusinov (YSR) subgap states realized in a double quantum dot. 
This sub-gap transport relies on intrinsic quasiparticle relaxation, but the tunability of the device allows us to explore also an additional relaxation mechanism based on charge transferring Andreev reflections. The transition between these two relaxation regimes is identified in the experiment as a marked gate-induced stepwise change in conductance.
We present a transport calculation, including YSR bound states and multiple Andreev reflections alongside with quasiparticle relaxation, due to a weak tunnel coupling to a nearby normal metal, and obtain excellent agreement with the data.
\end{abstract}

\maketitle

Superconductors are characterized by the existence of a Cooper-pair condensate with quasiparticle excitations, appearing above the superconducting gap, $\Delta$. The interplay between superconductivity and various types of impurities~\cite{Balatsky2006May, Yazdani1767, Franke2011May, Chatzopoulos2021Jan}, junctions~\cite{Beenakker1991Jun, Tosi2019Jan} and barriers~\cite{Prada2020Oct, Nichele2017Sep} can lead to the formation of localized quasiparticle states with energies smaller than the superconducting gap. Such subgap bound states are receiving increasing attention, as the parity protection offered by the gap makes them amenable to quantum coherent manipulation~\cite{Janvier2015Sep, Larsen2015Sep, Hays2018Jul, Hays2021Jan}. This attribute makes subgap states excellent candidates for qubits in quantum information processing. 

Nevertheless, many experiments have shown the existence of quasiparticle relaxation and poisoning, which break parity and decohere the subgap states~\cite{Tosi2019Jan, Hays2018Jul, Zgirski2011Jun}. The physics behind relaxation and poisoning processes differs from system to system as it depends on fabrication details and on the electromagnetic environment. It is therefore a priori difficult to estimate its origin and magnitude~\cite{Olivares2014Mar, Wang2014Dec, Marin-Suarez2020Jul}. The transport properties of subgap states depend strongly on the relaxation and poisoning rates and can therefore be used to probe the subgap states population dynamics~\cite{Martin2014, Ruby2015}. In a recent paper~\cite{Huang2020Dec}, direct transport between two sets of YSR states induced by magnetic impurities
was observed in a scanning tunnelling microscopy setup, yielding a clear measure of subgap dynamics independent of temperature and environmental broadening.

In this Letter, we investigate the subgap dynamics in a setup based on Coulomb blockaded quantum dots coupled to superconductors, which provides \textit{continuous} tunability of the subgap state energies. This tunability gives us unprecedented access to the full phase diagram of relaxation processes within a single device.
\begin{figure}[H]\includegraphics[width=1\linewidth]{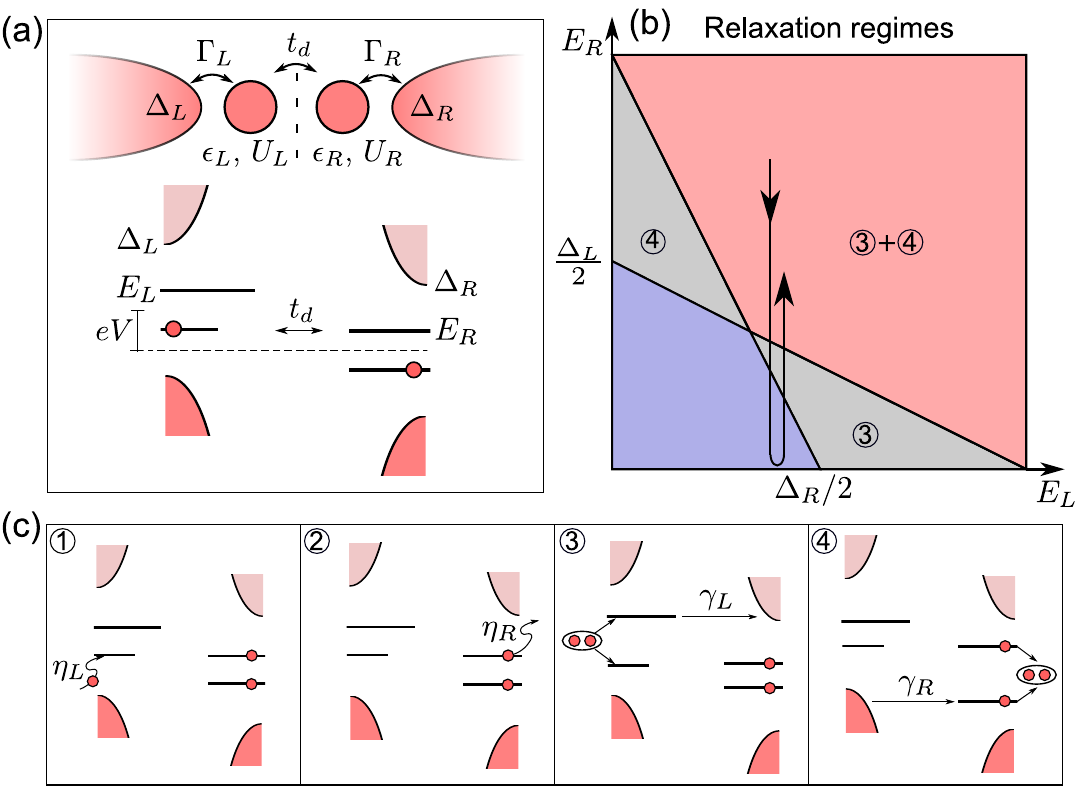}
\caption{(a) Schematic of the S-DQD-S system and energy level diagram for direct bound state to bound state transport at $eV = E_L+E_R$. System shown in the $\ket{0,0}$ state. (b) Bound state energy map delineating regions with different relaxation processes, indicated by numbers and illustrated in (c). Processes 1 and 2 are available in all sectors of the diagram. The indicated path corresponds to the change in relaxation regimes seen in Fig.\ref{fig2}(c) as the plunger gate voltage of the right dot is swept. (c) Four different relaxation mechanisms available at different subgap state energies. Processes 1 and 3 (2 and 4) relate to the left (right) quantum dot and are all depicted starting from the doubly excited state $\ket{1,1}\rightarrow \ket{0,1}$ ($\ket{0,1}$). Processes 1 and 2 refer to intrinsic quasiparticle relaxation, while 3 and 4 employ Andreev reflection via the continuum of the opposite side and transfers a net charge.
}
\label{fig1}
\end{figure}
To explain the transport signatures, we utilize Floquet Keldysh Green functions~\cite{Cuevas1996Sep, Villas2020Jun} to calculate the current across different relaxation regimes, and we demonstrate that these results can be understood in terms of master equations as in Ref.~\onlinecite{Huang2020Dec}, which we extend here to all relaxation regimes.

The interaction between a superconductor and the spin localized on a quantum dot leads to the formation of a YSR state~\cite{Jellinggard2016, Lee2014Jan, Pillet2010Dec}. The quantum dot is characterized by a charging energy, $U$, a level position, $\epsilon$, and a tunnel coupling to a superconductor characterized by a tunnelling rate $\Gamma$. By tuning a gate voltage to change $\epsilon$, one can manipulate both the excitation energy and the ground state of the superconductor-dot system. We use an InAs nanowire based double quantum dot (DQD) coupled to two superconductors~\cite{Saldana2018, Su2017Sep, Bouman2020Dec} to obtain two independent subgap states at energies $E_L$ and $E_R$, shown schematically in Fig.~\ref{fig1}(a). 

In the limit of low tunnel coupling between the dots, $t_d$, compared to the dot-superconductor tunnelling rates, $\Gamma_L$ and $\Gamma_R$, each dot will be in equilibrium with its respective superconductor, and a bias voltage, $V$, applied across the superconductors will cause a voltage drop across the two dots. At the resonances, $eV = \pm (E_L + E_R)$, the electron component of one subgap state is aligned with the hole component of the other, and direct electron transfer can take place. This will excite both subgap systems, i.e. $\ket{0,0}\leftrightarrow\ket{1,1}$, where $0(1)$ denote the ground(excited) state in the corresponding left, or right subgap system. The potential for such resonant transitions to carry a current relies entirely on the availability of relaxation channels to reset the subgap systems back to $\ket{0,0}$ after each inter-dot tunnelling process.

A diagram showing the different relaxation regimes and a schematic of available relaxation processes are presented in Figs.~\ref{fig1} (b) and (c). The intrinsic relaxation processes 1 and 2, with rates $\eta_{L/R}$, in Fig.\ref{fig1} (c) are active at all energies, while processes 3 and 4, with rates $\gamma_{L/R}$, only become available for subgap states with $E_{R/L} + 2E_{L/R} > \Delta_{R/L}$, where an Andreev reflection, followed by a single quasiparticle transfer to the opposing continuum may serve to reset the subgap excitations. Since these additional relaxation channels themselves transfer charge, a full transport cycle using both processes 3 and 4 constitutes a transfer of three electrons in total. Notice that, unlike multiple Andreev reflection (MAR) processes between two superconductors~\cite{Averin1995Aug}, this 3-electron transfer occurs incoherently. In total, one should therefore expect a higher relaxational current through electron/hole-aligned subgap states when the $eV=E_L+E_R$ resonance occurs above the threshold bias, i.e. for $|eV|>\min
(|\Delta_{L}-E_{R}|,|\Delta_{R}-E_{L}|)$ corresponding to the grey and red regions of Fig.~\ref{fig1}(b).

\begin{figure}[t]\includegraphics[width=1\linewidth]{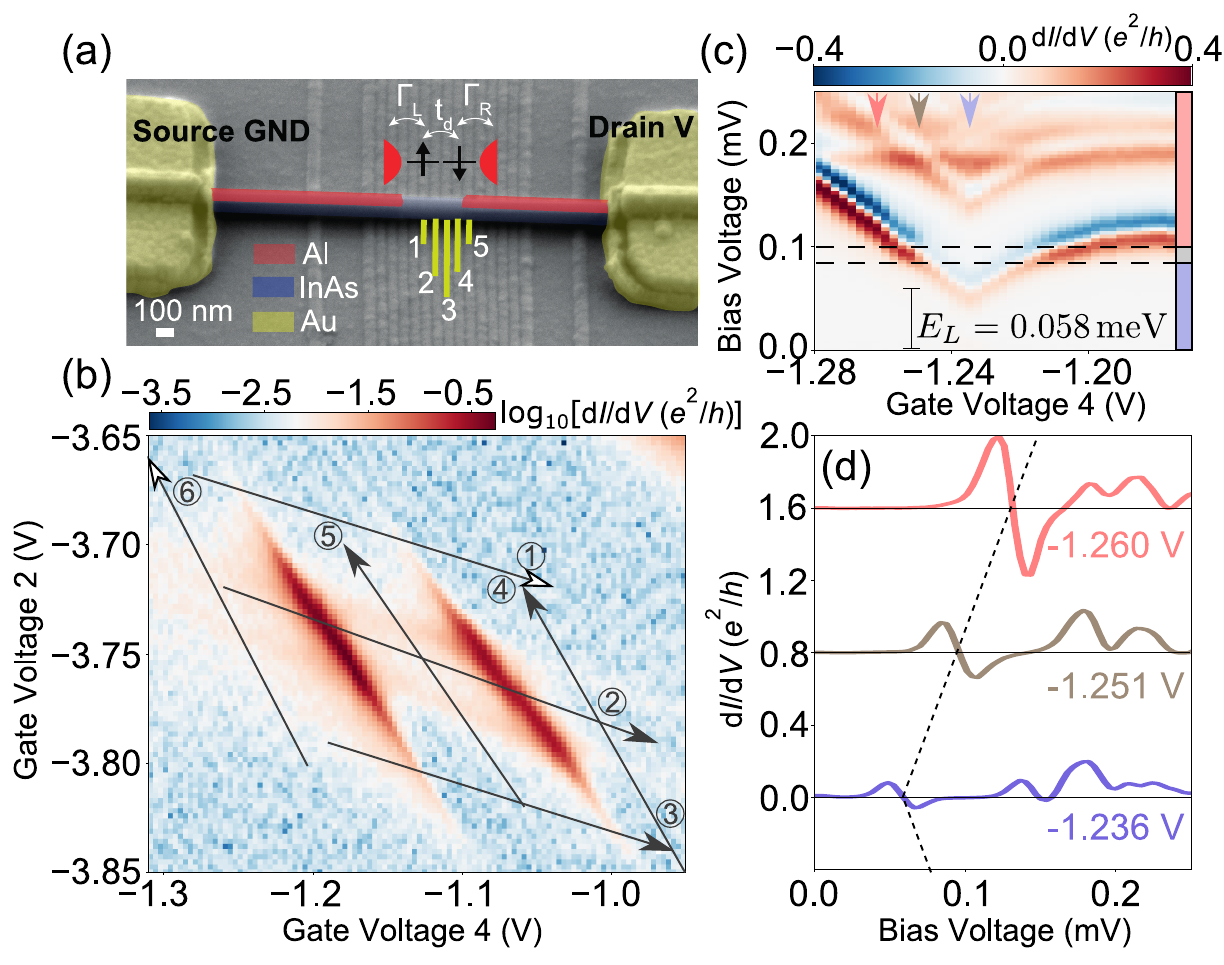}
\caption{(a) False colored scanning tunneling micrograph of the device. A schematic of the dots is shown at the junction. (b) Zero-bias conductance charge diagram in logarithmic scale. Arrows indicate linecuts plotted in (c) and in Fig.~\ref{fig3}. Gates 1, 3, 5 are set to -9.05 V, -8.7 V, and 0.74 V, respectively, and the backgate is set to 11.15 V. (c) Conductance as a function of bias, and gate voltages following half the range of linecut 1 in (a) parametrized by gate 4. A vertical colorbar on the right indicates relaxation regimes for $E_L = 0.058$ meV (read off as indicated) with colors indicating the corresponding regime in Fig.~\ref{fig1}(b). Arrows in the top mark cuts shown in (d) and dashed horizontal lines indicate changes in relaxation regimes. (d) Conductance vs. bias voltage along three vertical cuts in (c) placing the $eV=E_L+E_R$ resonance in different relaxation regimes, as indicated by color. Each cut is vertically displaced by 0.8$e^2/h$. The dotted line traces the movement of the resonance. 
}
\label{fig2}
\end{figure}
{\it Measurements} are carried out in a device investigated earlier at different gate settings in Refs.~\onlinecite{Saldana2018, EstradaSaldana2020Nov}, based on a 110 nm-diameter InAs nanowire with 7 nm superconducting aluminum grown in-situ epitaxially on three facets of the wire. The wire is deposited on top of an array of gates insulated by 20 nm of hafnium oxide, which is used to define the double dot architecture, and contacted by Ti/Au leads on each side. Aluminum is etched away before contact deposition to form a 350 nm long junction. The device is equipped with a global Si/SiOx substrate backgate. A scanning tunneling micrograph of the device is shown in Fig.~\ref{fig2}(a). Gates 1, 3 and 5 control the tunnel couplings, $\Gamma_L$, $t_d$ and $\Gamma_R$, and are set to constant voltages. Plunger gates 2 and 4 control the filling of the corresponding left and right dots. 

This device and its connecting circuitry has been characterized in Refs.~\onlinecite{Saldana2018, EstradaSaldana2020Nov}, where it was tuned up to measure (critical) supercurrent for different regimes of YSR screening. In this work, the device is adjusted differently to explore the relaxational bound-state-to-bound-state currents illustrated in Fig.~\ref{fig1}. To this end, we scan zero-bias conductance using standard lock-in measurements and locate a shell with no apparent anti-cross between charge sectors, indicative of weak inter-dot tunnel coupling, $t_d$, and charging energy, $U_d$. A logarithmic map of the conductance is shown in Fig.~\ref{fig2} (b). Pairs of vertical and horizontal stripes in the map are due to a combination of supercurrent and subgap resonances crossing zero energy~\cite{EstradaSaldana2020Nov,SupMat}. From independent measurements~\cite{SupMat}, we find $U_L, \hspace{0.05cm} U_R\approx 2$ meV and $\Delta_L, \hspace{0.05cm}\Delta_R \approx 0.14$ meV consistent with a YSR interpretation of subgap states.

\begin{figure}[t]\includegraphics[width=\linewidth]{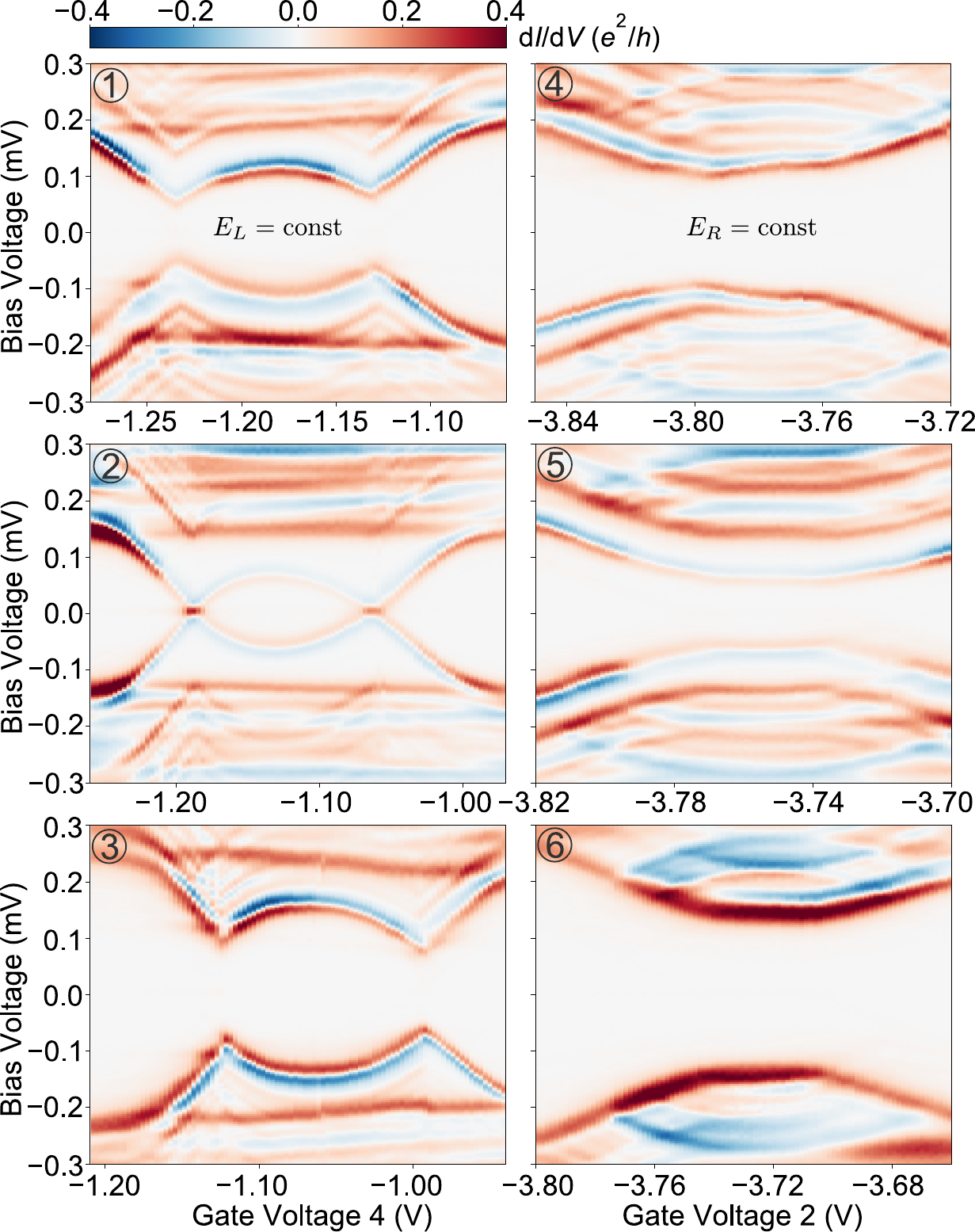}
\caption{Experimental data showing conductance as a function of gate, and bias voltage for the six line cuts shown in Fig.~\ref{fig2}(b). Cuts 1-3 (4-6) are parametrized by gate 4 (2), but gate 2 (4) is also tuned for each cut to fullow the lines indicated in Fig.~\ref{fig2}(b). Cuts 1-3 are horizontal cuts tuning the right dot, while cuts 4-6 are vertical cuts tuning the left dot. All plots are made with the same color scale.}
\label{fig3}
\end{figure}
In Fig.~\ref{fig2}(c), we show half of the gate extension of the central line cut of differential conductance versus bias and gate voltages labeled 2 in Fig.~\ref{fig2}(b). To interpret this cut, we assume that the energy of the left subgap state, $E_L$, remains constant as the right dot is gated, and identify the lowest lying feature as the $eV=E_L+E_R$ resonance, supported by the negative differential conductance (NDC) immediately following the conductance peak. As gate 4 is tuned, a sudden change of slope occurs at $-1.236$ V, which indicates that $E_R=0$, signalling a change of ground state of the right dot-superconductor system, and allows us to infer that $E_L = 0.058$ meV$<\Delta_R/2$.

Strikingly, as the $eV = E_L + E_R$ feature in Fig.~\ref{fig2}(c) moves with gate 4, stepwise changes in conductance are observed before and after the phase transition. The position of these thresholds fits with changes in the available relaxation processes, estimated from the bound-state energies, shown as horizontal lines in Fig.~\ref{fig2}(c) and as the path in Fig.~\ref{fig1}(b). This path shows that as gate increases the resonance moves from red$\rightarrow$blue$\rightarrow$red with grey regions only observed as transitional steps. In Fig.~\ref{fig2}(d), three linecuts show the decrease in conductance of the lowest lying peak-dip features by approximately a factor of 4 between the top, and bottom curves. This pronounced contrast in conductance marks a gate tunable transition between three different relaxation regimes.

\begin{figure}[t]\includegraphics[width=\linewidth]{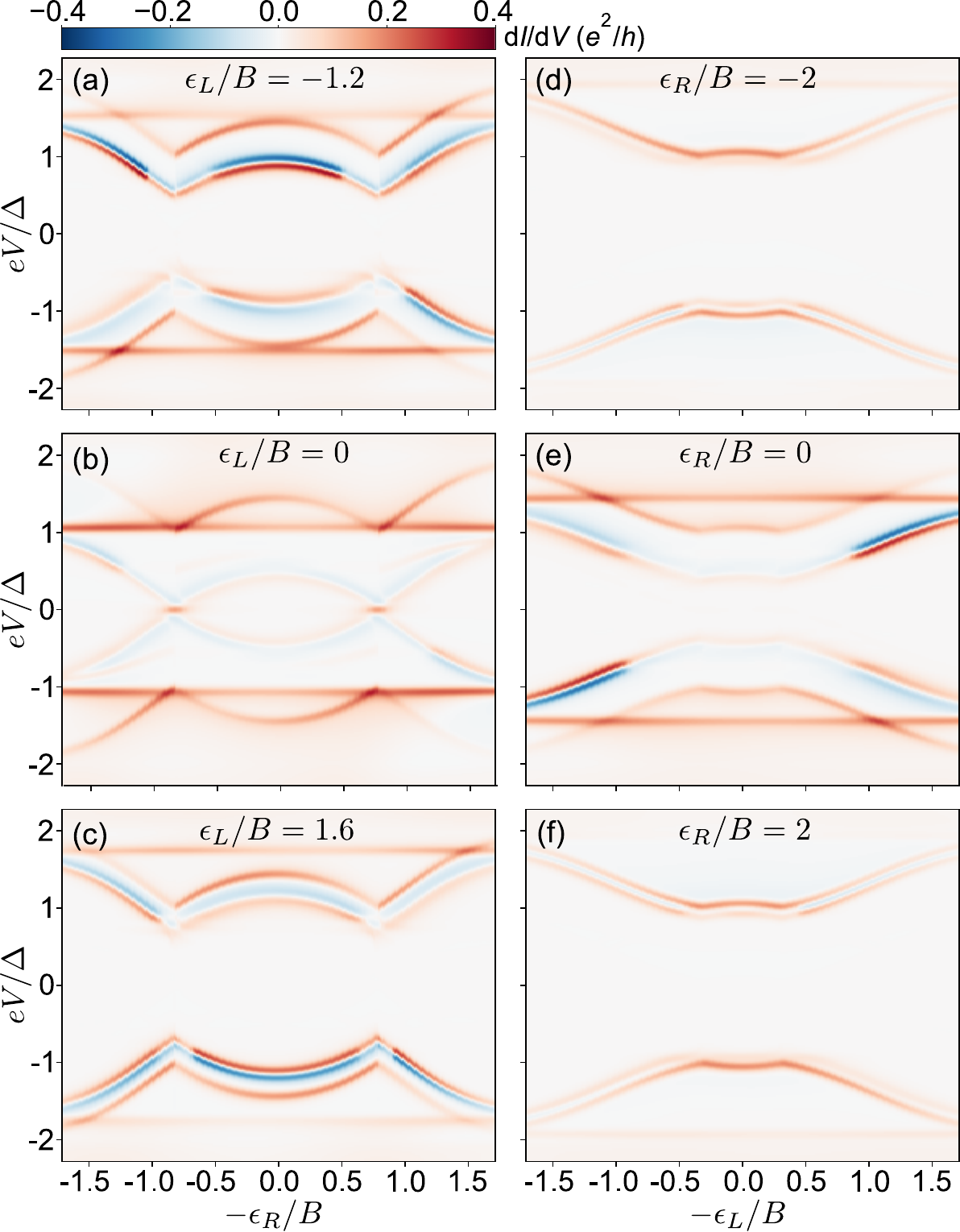}
\caption{Conductance as a function of normalized gate, and bias voltage for six linecuts, calculated using Keldysh Floquet Green functions~\cite{SupMat}. $\epsilon_{L/R}$ are chosen so as to match the cut in the corresponding panel in Fig.~\ref{fig3}. All plots are made with the parameters listed in~\cite{Parameters}, using the same color scale as in Fig.~\ref{fig3}.
}
\label{fig4}
\end{figure}
These types of changes in conductance at special thresholds are widespread in our data and their positions match expectations from Fig.~\ref{fig1}(b). In Fig.~\ref{fig3} we plot the 6 linecuts indicated in Fig.~\ref{fig2}(b), where the lowest lying feature corresponds to $eV = E_L + E_R$. From the slope of this feature we infer that the right dot is intermediately coupled to the superconductor showing a characteristic eye shape, while the left dot is more strongly coupled and close to the phase transition at the particle-hole symmetric point~\cite{Bauer2007Nov}.

Additional conductance features at higher bias in Fig.~\ref{fig2}(c) are identified as a peak at $eV = E_{R}+\Delta_{L}$ dispersing like the $eV = E_L+E_R$ feature, and a peak at $eV = E_{L}+\Delta_{R}$, which is independent of gate 4, supporting that $E_L$ remains constant as $E_R$ is tuned by gate 4.
In all cuts shown in Fig.~\ref{fig3}, replicas of the $eV = E_{L/R}+\Delta_{R/L}$ features are seen above the first such feature. In cuts 1-3, these appear as repetitions of the $E_{L/R}+\Delta_{R/L}$ features, while in cuts 4-6 features with the opposite slope of the subgap state also appear. Similar features have been observed in other devices~\cite{Su2018Sep, Whiticar2021Jan} and we ascribe them to multiple sub-bands in the proximitized InAs nanowire~\cite{EstradaSaldana2020Nov}. In this scenario, a conductance peak would appear for each sub-band coherence peak as the bias voltage is increased~\cite{SupMat}. 

We model the DQD as two Anderson models with superconducting leads and an additional interdot tunnel coupling. For simplicity, we employ a spin-polarized mean field approximation~\cite{Villas2020Jun, Villas2021Apr}, which is known to capture the characteristic gate-dependence of the YSR state~\cite{Zonda2015Mar, Pillet2010Dec}.
This artificially spin-polarized description omits inter-dot exchange, which is anyway negligible since $t_d\ll U_{L/R}$. 
To circumvent artificial spin-blockade, the spin-polarizing mean fields are chosen to point in orthogonal directions on each dot: ${\boldsymbol B}_{L}={\boldsymbol {\hat z}}U_{L}/2$ and ${\boldsymbol B}_{R}={\boldsymbol {\hat x}}U_{R}/2$ \cite{Villas2021Apr}. With these caveats, we regard the model as a qualitative description of the experimental situation. 

To calculate the nonlinear $I-V$ characteristics, we employ Keldysh Floquet Green functions incorporating both MAR, and relaxation processes. The current is $P(E)$-broadened by a  Gaussian of width $\sigma=0.04\Delta\approx 6\, \mu$eV before calculating the conductance~\cite{SupMat}. 
Results of the calculations are shown in Fig.~\ref{fig4}. Parameters are kept fixed except for $\epsilon_L$ and $\epsilon_R$, which are chosen so as to match the linecuts shown in Fig.~\ref{fig3}. Tunnelling rates, $\Gamma_{L/R}$, are chosen such that the slope with gate voltage of each YSR state matches the experiment. Intrinsic relaxation rates are assumed symmetric, $\eta_L=\eta_R$, and together with $t_d$ they are tuned to match the overall conductance scale and the size of conductance steps between different relaxation regimes. 
In the calculations shown in Fig.~\ref{fig4}, we observe the previously described $eV=E_L+E_R$ and $eV=\Delta_{L/R}+E_{R/L}$ features alongside the stepwise changes in conductance at transitions between different relaxation regimes. 

Some analytical insight on the relaxational current carried at $eV = E_L+E_R$ can be obtained by solving a phenomenological master equation of the Lindblad form~\cite{Breuer}
As detailed in the supplement, this leads to a Lorentzian current peak,
\begin{align} \hspace*{-0.2cm}
I = \frac{e}{h} \frac{2\pi\gamma_e^2\left[ \Lambda_L\left(1+\frac{\gamma_R}{\Lambda_R}\right)+\Lambda_R\left(1+\frac{\gamma_L}{\Lambda_L}\right)\right]}{\gamma_e^2\frac{\left(\Lambda_L+\Lambda_R\right)^2}{\Lambda_L\Lambda_R}+\frac{\left(\Lambda_L+\Lambda_R\right)^2}{4}+(eV-E_L-E_R)^2}, \label{MasterEq}
\end{align}
where $\gamma_e^2 = v_L^2u_R^2 t_d^2$ is the rate of electron transfer between the left hole component with amplitude $v_L$, and the right electron component with amplitude $u_R$. The total relaxation rate for each side is $\Lambda_{L/R}=\eta_{L/R}+\gamma_{L/R}$ with $\eta_{L/R}$ being the intrinsic relaxation rate, and $\gamma_{L/R}$ the rate of relaxation occurring via Andreev reflections as sketched in Fig.~\ref{fig1}(c). Using Fermi's golden rule, we infer the rates to be $\gamma_L=\pi u_L^2 t_d^2 d_R(2E_L+E_R)$ and $\gamma_R=\pi v_R^2 t_d^2 d_L(-2E_R-E_L)$ with $u_L$ ($v_R$) being the corresponding electron (hole) component amplitudes and $d_{L/R}(E)$ the density of states at energy $E$. For the corresponding $eV = -E_L-E_R$ peak let $E_{L/R} \rightarrow -E_{L/R}$, substitute $u$ and $v$, and the above formulas apply. As shown in the Supplemental Material, these formulas perfectly match the results obtained from Keldysh Floquet Green functions for $eV=\pm(E_L+E_R)$. In the limit $\eta_{L/R}\gg\gamma_{L/R}, \gamma_e$, Eq.~\eqref{MasterEq} reduces to Fermi's golden rule, and the bias asymmetry reflects directly the ratio between electron and hole amplitudes, $u_{R}^{2}v_{L}^{2}/v_{R}^{2}u_{L}^{2}$. For $\eta_{L}=\eta_{R}$ and $\gamma_L=\gamma_R=0$, which is the regime relevant in the blue region of Fig.~\ref{fig1}(b), Eq.~\eqref{MasterEq} reproduces the results of Ref.~\cite{Huang2020Dec}. 
In the regime relevant for the present experiment, $t_d\gg \eta_{L/R}$ and hence $\gamma_e,\gamma_{L/R}\gg \eta_{L/R}$ when outside of the blue region in Fig.~\ref{fig1}(b), the bias asymmetry appears reversed compared to the Fermi's golden rule limit~\cite{SupMat}. Comparing Figs.~\ref{fig3}~and~\ref{fig4}~(a, e), this asymmetry is seen to be reproduced by the transport calculation. A similar reversed asymmetry has been observed also by STM spectroscopy of YSR states probed by a superconducting continuum at $eV = E_{L/R}+\Delta_{R/L}$~\cite{Ruby2015}.

Extending the master equation to include the doublet nature of the odd-parity subgap states, we find that the relaxational current generally depends on the ground state (odd-parity doublet or even-parity singlet), and that a finite spin relaxation rate, $\Gamma_s$, must be included in order to avoid spin-blockade. Such spin relaxation has been measured in a similar device~\cite{Hays2021Jan}. Consistency with the experimental data requires that $\gamma_{e}\gg\Gamma_s \gg \eta_{L,R}$~\cite{SupMat}. 

Without independent estimates of $t_d$, $\Gamma_s$ and the continuum density of states $d_{L/R}(E)$, we cannot confidently extract intrinsic relaxation rates $\eta_{L,R}$. Nevertheless, a number of qualitative conclusions can be drawn: 1) We observe only very weak subgap mirages~\cite{Su2018Sep, Kumar2014Feb} indicative of a hard gap~\cite{SupMat}; 2) Intrinsic relaxation must be present and be largely independent of the bound-state energy; 3) No quasiparticle poisoning, spontaneously exciting the ground state, is observed, since this would lead to lines at $eV = E_{L}-E_{R}$~\cite{Huang2020Dec} and $eV =\Delta_{L/R}-E_{R/L}$~\cite{Kumar2014Feb} with opposite gate-voltage curvature. The last two observations indicate that the intrinsic relaxation is neither due to quasiparticle poisoning in the leads nor to high-energy phonon/photon modes~\cite{Olivares2014Mar}. More likely, the nearby normal metallic Ti/Au leads act as quasiparticle traps with a weak tunnel coupling to the quantum dots. This is consistent with our modelling of subgap-state relaxation as arising from a weak tunnel coupling to a large-bandwidth metallic lead, which also explains the weak low-voltage mirages observed in the experiment~\cite{SupMat}.

In conclusion, we have presented measurements of direct transport between two subgap states in a DQD setup. The electrical tunability of this setup allowed us to explore the transition between two different relaxation regimes, identified as stepwise changes in conductance along the $eV=E_L +E_R$ subgap resonance. We developed a model for the gateable subgap states, including intrinsic relaxation via weak tunnel coupling to a nearby normal metal, and a transport calculation combining MAR and relaxation was found to explain the observed signatures and provided excellent agreement with the experimental data. 
The presented bound-state-to-bound-state measurements hinge on the availability of intrinsic relaxation processes, yielding key insights into the underlying population dynamics of gateable subgap states relevant for future designs of superconducting qubits.

\begin{acknowledgments}
The authors thank Juan Carlos Cuevas and Christian Ast for fruitful discussion. The project received funding from the European Union’s Horizon 2020 research and innovation program under the Marie Sklodowska-Curie grant agreement No.~832645. We additionally acknowledge financial support from the Carlsberg Foundation, the Independent Research Fund Denmark, QuantERA ’SuperTop’ (NN 127900), the Danish National Research Foundation, Villum Foundation project No.~25310, and the  Sino-Danish Center. P.~K. acknowledges support from Microsoft and the ERC starting Grant No.~716655 under the Horizon 2020 program. J.N., K.G-R and A.L.Y. acknowledge European Union’s Horizon 2020 research 
and innovation programme for financial support grant No.~828948 (AndQC).
A.L.Y. acknowledges support by Spanish MICINN through grants 
FIS2017-84860-R and through the “Mar\'{\i}a de Maeztu” Programme for 
Units of Excellence in R\&D (Grant No.~MDM-2014-0377).
\end{acknowledgments}

\bibliographystyle{apsrev4-1}
\bibliography{Bibtex}

\begin{thebibliography}{41}
\bibitem [{Par()}]{Parameters}%
  \BibitemOpen
  \href@noop {} {}\bibinfo {note} {Parameters used in plots:
  $\Delta_L=\Delta_R=\Delta=0.14$meV, $B_L=B_R = 7$, $\Gamma_L = 6.65$,
  $\Gamma_R=4.2$, $t_d = 0.73$, $\eta_L=\eta_R=3.75\times 10^{-4}$, $T=10^{-3}$
  and $\sigma = 0.04$ all in units of $\Delta$. The BCS density of states of
  the superconducting leads are broadened by $\eta_s=10^{-6}\Delta$ for
  numerical reasons.}\BibitemShut {Stop}%
\end{thebibliography}

\end{document}